\documentclass[]{raa} 

\usepackage{graphicx,times}  
\usepackage{newtxtext,newtxmath}
\usepackage[T1]{fontenc}
\usepackage{ae,aecompl}
\usepackage{amsmath}
\usepackage{amssymb}

\def\kms{$\mathrm{km\,s}^{-1}$}
\def\ms{$\mathrm{m\,s}^{-1}$}

\begin{document}

\title{The search for roAp stars: null results and new candidates from Str{\"o}mgren-Crawford photometry}

   \volnopage{Vol.0 (20xx) No.0, 000--000}      %%preserved for Editor. DOn't remove!
   \setcounter{page}{1}          %%starting page, preserved for Editor. DOn't remove!

\author{E. Paunzen
   \inst{1}
\and G. Handler
   \inst{2}
\and K. Ho{\v n}kov{\'a}
   \inst{3}
\and J. Jury{\v s}ek
   \inst{3}
\and M. Ma{\v s}ek
   \inst{4}
\and M. Dr{\'o}{\.z}d{\.z}
   \inst{5}
\and J. Jan{\'i}k
   \inst{1}
\and W. Og{\l}oza
   \inst{5}
\and L. Hermansson
   \inst{6}
\and M. Johansson
   \inst{6}
\and M. Jel\'{i}nek
   \inst{7}
\and M. Skarka
   \inst{7}
\and M. Zejda
   \inst{1}
}

% List of institutions
\institute{Department of Theoretical Physics and Astrophysics, Masaryk University,
Kotl\'a\v{r}sk\'a 2, CZ-611\,37, Czech Republic; {\it epaunzen@physics.muni.cz} \\
\and Nicolaus Copernicus Astronomical Center, Bartycka 18, 00-716 Warsaw, Poland\\
\and Variable Star and Exoplanet Section of Czech Astronomical Society, Vset{\'i}nsk{\'a} 
941/78, CZ-757\,01, Vala{\v s}sk{\'e} Mezi{\v r}{\'i}{\v c}{\'i}, Czech Republic\\
\and Institute of Physics, The Czech Academy of Sciences, Na Slovance 1999/2, CZ-182\,21, Praha, Czech Republic\\
\and Mt. Suhora Observatory, Pedagogical University, Podchor\c a\. zych 2, 30-084 Krak{\'o}w, Poland\\
\and Sandvretens Observatory, Linn{\'e}gatan 5A, 75332, Uppsala, Sweden\\
\and Astronomical Institute, Czech Academy of Sciences, (ASU CAS), Ond\v{r}ejov, Czech Republic\\
}

   \date{Received~~2018 month day; accepted~~2018~~month day}
	
\abstract{ 
The rapidly oscillating Ap (roAp) stars exhibit pulsational 
photometric and/or radial velocity variations on time scales of several 
minutes, which are essential to test current pulsation models as well as 
our assumptions of atmospheric structure characteristics. In addition, 
their chemical peculiarity makes them very interesting for probing 
stellar formation and evolution in the presence of a global magnetic 
field. To date, a limited number of only 61 roAp stars are known to 
show photometric variability. On the other hand, a literature survey 
yields 619 unique stars that have unsuccessfully been searched for 
variability of this kind. Str{\"o}mgren-Crawford uvbybeta 
photometry of stars from both subgroups were used to investigate whether there is 
a selection bias of the investigated stars. We also present new photometric measurements (202 hours 
in 59 different nights) of 55 roAp candidates. We did not detect any new 
roAp star. Although our detection limits are comparable to other surveys, we also 
did not find pulsations in the known roAp star HD 12098, which may be a consequence 
of temporal amplitude changes. On the other hand, we do find some evidence for 
photometric variability of beta CrB at its spectroscopically derived pulsation period.
From the uvbybeta photometry we conclude that the blue border of the roAp 
instability strip appears observationally well defined, whereas the red
border is rather poorly known and studied. Within these boundaries,
a total of 4646 candidates were identified which appear worthwhile to be 
investigated for short-term pulsational variability.
}

   \authorrunning{E. Paunzen et al. }            %author_head in even pages
   \titlerunning{Null results and new candidates from Str{\"o}mgren-Crawford photometry }  % title_head in odd pages

   \maketitle
	
\section{Introduction}

Up to now, there are 61 rapidly oscillating Ap (roAp) 
stars known (Joshi et al.~\cite{Josh16}). They are located within an area
of pulsational instability in the Hertzsprung-Russell diagram,
across the main sequence, ranging in effective temperature 
from about 6\,600\,K to 8\,500\,K. The roAp stars share this area
with pulsators of the $\delta$ Sct and $\gamma$ Dor types, but their
spectral and variability properties are markedly different.

Photometric investigations of roAp stars show a period 
range of five to twenty five minutes, which is consistent
with acoustic (p-mode) pulsations of low degree and high 
radial overtone (Kurtz et al.~\cite{Kurt11}).

The driving mechanism of their oscillation modes is still not fully 
understood. The most probable explanation appears to be the 
``classical'' $\kappa$-mechanism operating in the hydrogen ionisation 
zone (e.g. Balmforth et al.~\cite{balmforth01}). Many physical processes could play a 
role in this context: for example the coupling with the magnetic field, 
the ability of the latter to freeze convection and to facilitate the 
stratification of chemical elements. Therefore, the roAp stars are very 
interesting for studying pulsations in the presence of a stable 
magnetic field and diffusion in the stellar atmosphere. The very small 
percentage of such variables with respect to the total number of Ap 
stars already suggests that the conditions needed to excite pulsation 
with a detectable amplitude have to be very specific.

Even so, there appear to be additional shortcomings in the 
theoretical prediction of pulsation among roAp stars. For instance, 
theoretically predicted regions of pulsational instability 
(Cunha~\cite{cunha02}) do not agree with the observed ones, in the sense that 
both the theoretical blue and red edges are shifted towards higher 
temperatures compared to observations. Furthermore, some roAp stars 
pulsate with frequencies higher than the acoustic cutoff frequency, 
which led to the suggestion that these could be driven by turbulent 
pressure (Cunha et al.~\cite{cunha13}).

Therefore, as important as the knowledge about the roAp stars 
themselves, is the analysis of objects in the same temperature range 
which show no sign of roAp variability (noAp stars). A comparison of 
both groups should provide crucial constraints about the astrophysical 
conditions that enable pulsations in those stars.

We present in this paper new photoelectric and CCD photometric 
observations for 55 roAp candidates selected on the basis of their 
spectral types. 
In total, we analysed 202 hours of 
observations in 59 different nights. No new roAp star was detected.

\begin{table*}
\begin{center}
\caption[]{Observation log, and the results from the time series analysis. HD 12098 and HD 137909
are known roAp stars.}
\label{log}
\begin{tabular}{cccccccc}
\hline\noalign{\smallskip}
HD/TYC & BD/HIP/TYC & HJD(start) & $\Delta t$ & $N$ & Filter & Site & Upper Limit \\
 & & [2450000+] & [min] & & & & [mmag] \\
\hline\noalign{\smallskip}
8441 & 6560 & 6976.50705 & 113.5 & 327 & $B$ & Suhora (gs8in) & 0.7 \\
         & & 7012.28702 & 104.1 & 173 & $B$ & Uppsala & 5.8 \\
10783 & 10783 & 6977.18299 & 261.7 & 983 & $B$ & Suhora (gs8in) & 0.4 \\
12098 & 3687--2649--1 & 6976.53518 & 85.3 & 457 & $v$ & Suhora (Zeiss) & 0.3 \\
				 & & 6977.34519 & 21.7 & 99 & $v$ & Suhora (Zeiss) & 0.6 \\
12288 & 9604 & 6977.17831 & 258.7 & 1289 & $v$ & Suhora (Zeiss) & 0.4 \\
15089 & 11569 & 6977.27179 & 336.5 & 239 & $B$ & Suhora (MK) & 1.4 \\
         & & 6977.27204 & 337.5 & 443 & $B$ & Suhora (MK) & 1.8 \\
				 & & 7012.40129 & 90.7 & 161 & $B$ & Uppsala & 1.8 \\
20476 & 2852--1042--1 & 7083.32199 & 74.0 & 131 & $g'$ & BOOTES-2 & 3.4 \\
21190 & 14700 & 7053.67151 & 87.3 & 57 & $B$ & FRAM-Auger & 8.7 \\ 
25354 & 18912 & 7019.33714 & 162.5 & 292 & $B$ & Uppsala & 1.3 \\
38823 & 27423 & 7045.53101 & 78.9 & 65 & $B$ & FRAM-Auger & 5.8 \\
         & & 7059.58522 & 88.8 & 90 & $B$ & FRAM-Auger & 5.9 \\
47074 & 146--1489--1 & 7059.42474 & 133.7 & 267 & $g'$ & BOOTES-2 & 1.4 \\
50403 & 1343--1516--1 & 7051.50925 & 145.7 & 215 & $g'$ & BOOTES-2 & 7.0 \\
51684 & 33375 & 7045.70282 & 90.1 & 54 & $B$ & FRAM-Auger & 5.5 \\
         & & 7050.74948 & 124.3 & 96 & $B$ & FRAM-Auger & 4.8 \\
52181A & 1352--1045--1 & 7070.47641 & 147.9 & 252 & $g'$ & BOOTES-2 & 3.8 \\
52628 & 34174 & 7141.62517 & 68.6 & 147 & $B$ & APT & 0.8 \\
54824 & 34773 & 7140.62067 & 83.1 & 187 & $B$ & APT & 0.7 \\
         & & 7145.62726 & 50.6 & 127 & $B$ & APT & 0.7 \\
				 & & 7146.62815 & 49.6 & 94 & $B$ & APT & 2.9 \\
55228 & 1346--208--1 & 7143.62237 & 69.7 & 125 & $B$ & APT & 1.1 \\
61600 & 37483 & 6976.65123 & 77.1 & 388 & $v$ & Suhora (Zeiss) & 0.5 \\
65339 & 39261 & 6976.64251 & 83.5 & 334 & $B$ & Suhora (gs8in) & 0.5 \\
         & & 7014.99839 & 88.7 & 224 & $v$ & APT & 1.6 \\
				 & & 7015.99486 & 88.5 & 224 & $v$ & APT & 1.1 \\
				 & & 7019.48691 & 102.4 & 211 & $B$ & Uppsala & 4.1 \\
				 & & 7095.32740 & 86.4 & 194 & $B$ & Uppsala & 2.9 \\
68542 & 40311 & 7096.32184 & 126.2 & 172 & $B$ & Uppsala & 2.5 \\
         & & 7133.61644 & 81.5 & 192 & $B$ & APT & 0.8 \\
         & & 7134.61691 & 68.7 & 160 & $B$ & APT & 2.5 \\
         & & 7149.62957 & 62.6 & 151 & $B$ & APT & 1.2 \\
96003 & 54141 & 7109.73310 & 88.5 & 219 & $v$ & APT & 0.6 \\
         & & 7114.71982 & 62.9 & 151 & $v$ & APT & 0.6 \\
96097 & 54182 & 7143.67192 & 113.2 & 204 & $v$ & APT & 1.2 \\
         & & 7144.71328 & 43.9 & 109 & $v$ & APT & 1.4 \\
				 & & 7145.66500 & 112.2 & 278 & $v$ & APT & 1.3 \\
				 & & 7150.62999 & 164.5 & 416 & $v$ & APT & 2.6 \\
				 & & 7151.63058 & 164.6 & 416 & $v$ & APT & 2.2 \\
97633 & 54879 & 7092.77862 & 126.4 & 296 & $v$ & APT & 0.6 \\
         & & 7107.73745 & 88.2 & 219 & $v$ & APT & 0.6 \\
				 & & 7108.73344 & 88.1 & 219 & $v$ & APT & 0.7 \\
\noalign{\smallskip}\hline
\end{tabular}
\end{center}
\end{table*}
\addtocounter{table}{-1}
\begin{table*}
\begin{center}
\caption{continued.}
\begin{tabular}{cccccccc}
\hline\noalign{\smallskip}
HD/TYC & BD/HIP/TYC & HJD(start) & $\Delta t$ & $N$ & Filter & Site & Upper Limit \\
 & & [2450000+] & [min] & & & & [mmag] \\
\hline\noalign{\smallskip}
100809 & 56597 & 7457.77923 & 154.0 & 379 & $B$ & APT & 0.4 \\
          & & 7459.77421 & 154.0 & 381 & $B$ & APT & 1.0 \\
105999 & 59487 & 6735.76575 & 171.7 & 110 & $B$& FRAM-Auger & 1.0 \\
          & & 6736.53343 & 190.9 & 93 & $B$& FRAM-Auger & 1.3 \\
					& & 6736.53359 & 192.4 & 231 & $B$ & FRAM-Auger & 0.8 \\
107000 & 59998 & 7131.67513 & 126.5 & 311 & $B$ & APT & 0.6 \\
					& & 7133.67420 & 75.4 & 185 & $B$ & APT & 1.0 \\
107612 & 60313 & 7018.98447 & 114.1 & 280 & $v$ & APT & 0.7 \\
%         & & 7019.97737 & 125.1 & 300 & $v$ & APT & 111 \\
					& & 7020.98280 & 112.4 & 261 & $v$ & APT & 1.0 \\
					& & 7021.97194 & 125.1 & 246 & $v$ & APT & 0.6 \\
					& & 7095.43070 & 91.3 & 121 & $B$ & Uppsala & 3.6 \\
108283 & 60697 & 7109.79554 & 88.1 & 214 & $v$ & APT & 0.9 \\
          & & 7118.71067 & 124.0 & 274 & $v$ & APT & 0.8 \\
108945 & 61071 & 7019.98178 & 102.7 & 18 & $v$ & APT & 1.4 \\
          & & 7022.97093 & 115.4 & 20 & $v$ & APT & 1.6 \\
112528 & 63247 & 6739.81531 & 144.5 & 189 & $B$ & FRAM-Auger & 1.3 \\
          & & 6746.49682 & 67.2 & 32 & $B$ & FRAM-Auger & 9.8 \\
115708 & 64936 & 6709.59688 & 94.1 & 221 & $v$ & MUO & 0.6 \\
117361 & 65754 & 7092.86750 & 116.4 & 226 & $v$ & APT & 0.8 \\
          & & 7107.79969 & 88.4 & 217 & $v$ & APT & 0.7 \\
					& & 7108.79571 & 88.5 & 212 & $v$ & APT & 0.7 \\
118022 & 66200 & 7028.95317 & 123.0 & 276 & $v$ & APT & 0.8 \\
119213 & 66700 & 7025.96276 & 139.6 & 352 & $v$ & APT & 0.6 \\
124883 & 69661 & 7113.79238 & 88.7 & 224 & $v$ & APT & 1.1 \\
          & & 7118.81935 & 111.5 & 243 & $v$ & APT & 0.5 \\
          & & 7139.78190 & 75.8 & 192 & $v$ & APT & 1.3 \\
          & & 7148.79921 & 120.9 & 251 & $v$ & APT & 0.7 \\
125709 & 2011--105--1 & 7148.79789 & 133.9 & 30 & $v$ & APT & 2.9 \\
126515 & 70553 & 7113.72906 & 89.7 & 220 & $B$ & APT & 0.7 \\
          & & 7131.76541 & 140.0 & 285 & $v$ & APT & 2.9 \\
134305 & 74109 & 7140.68059 & 102.4 & 248 & $B$ & APT & 1.8 \\
          & & 7141.67654 & 102.9 & 253 & $B$ & APT & 1.0 \\
134793 & 74334 & 7140.75278 & 117.5 & 281 & $B$ & APT & 0.6 \\
          & & 7141.74905 & 115.3 & 288 & $B$ & APT & 0.5 \\
137909 & 75695 & 7092.95079 & 113.6 & 271 & $v$ & APT & 0.8 \\
          & & 7107.86215 & 88.1 & 221 & $v$ & APT & 0.9 \\
					& & 7108.85822 & 88.0 & 220 & $v$ & APT & 0.6 \\
          & & 7109.85830 & 88.0 & 219 & $v$ & APT & 0.9 \\
					& & 7135.95188 & 67.9 & 144 & $v$ & APT & 1.2 \\
          & & 7150.74538 & 215.7 & 544 & $v$ & APT & 1.3 \\
          & & 7151.74604 & 215.7 & 535 & $v$ & APT & 0.6 \\
242006 & 699--1159--1 & 7047.30371 & 96.8 & 260 & $g'$ & BOOTES-2 & 1.8 \\
\noalign{\smallskip}\hline
\end{tabular}
\end{center}
\end{table*}
\addtocounter{table}{-1}
\begin{table*}
\begin{center}
\caption{continued.}
\begin{tabular}{cccccccc}
\hline\noalign{\smallskip}
HD/TYC & BD/HIP/TYC & HJD(start) & $\Delta t$ & $N$ & Filter & Site & Upper Limit \\
 & & [2450000+] & [min] & & & & [mmag] \\
\hline\noalign{\smallskip}
242442 & 2390--208--1 & 7047.42182 & 98.6 & 190 & $g'$ & BOOTES-2 & 1.4 \\
245222 & 1869--1770--1 & 7063.43180 & 135.9 & 245 & $g'$ & BOOTES-2 & 2.8 \\
245416 & 2404--775--1 & 7045.53083 & 138.3 & 195 & $g'$ & BOOTES-2 & 2.1 \\
          & & 7048.49546 & 126.9 & 244 & $g'$ & BOOTES-2 & 5.6 \\
246587 & 1861--299--1 & 7046.44334 & 132.7 & 185 & $g'$ & BOOTES-2 & 2.1 \\
246726 & 2405--1110--1 & 7102.37786 & 122.9 & 711 & $B$ & Suhora (gs8in) & 1.5 \\
247628 & 2409--1922--1 & 7100.36153 & 159.6 & 740 & $B$ & Suhora (gs8in) & 1.5 \\
259380 & 2426--652--1 & 7072.48449 & 147.2 & 258 & $g'$ & BOOTES-2 & 2.2 \\
264291 & 1343--2532--1 & 7063.52746 & 88.3 & 160 & $g'$ & BOOTES-2 & 2.2 \\
266311 & 157--22--1 & 7059.51917 & 145.2 & 226 & $g'$ & BOOTES-2 & 2.3 \\
268471 & 1895--873--1 & 7047.49150 & 99.0 & 191 & $g'$ & BOOTES-2 & 2.1 \\
276625 & 2884--484--1 & 7050.30892 & 92.9 & 210 & $g'$ & BOOTES-2 & 2.4 \\
2320--605--1 & +36 363& 7036.19795 & 126.7 & 312 & $B$ & Suhora (gs8in) & 0.4 \\
3009--62--1 & +41 2144 & 7047.56523 & 86.6 & 224 & $g'$ & BOOTES-2 & 2.5 \\
3021--603--1 & +38 2360 & 7052.53427 & 95.5 & 228 & $g'$ & BOOTES-2 & 2.7 \\
4520--578--1 & 13172 & 7036.33613 & 97.5 & 262 & $B$ & Suhora (gs8in) & 4.0 \\
               & & 7045.43571 & 131.4 & 196 & $g'$ & BOOTES-2 & 2.4 \\
               & & 7048.28533 & 126.9 & 126 & $g'$ & BOOTES-2 & 3.5 \\
\noalign{\smallskip}\hline
\end{tabular}
\end{center}
\end{table*}

The classification of a star as a noAp object is not straightforward 
because roAp stars can show, for example, strong amplitude variations from 
night to night (e.g., White et al.~\cite{White11}; Medupe et al.~\cite{Medu15}). The amplitudes also 
depend on the wavelength region observed (Medupe \& Kurtz~\cite{Medu98}), and space 
photometry shows that some stars pulsate with amplitudes undetectable 
from the ground (e.g. Holdsworth~\cite{Hold16}). Therefore, a noAp star can only 
be defined by an observational upper limit of variability for a given 
time base and filter, which needs to be kept in mind when discussing 
these stars.

In order to select new candidates within the Str{\"o}mgren-Crawford 
$uvby\beta$ photometric system, we analysed the location of the roAp and 
noAp groups in reddening free diagrams. On the basis of newly defined 
areas in which roAp stars are located, we present a list of 4646 new 
candidates deemed worthwhile for variability searches. 
Several regions of Str{\"o}mgren-Crawford indices (for example,
$\lbrack m_1 \rbrack > 0.35$) appear to be insufficiently 
investigated for new variables.

\section{Observations and Data Reduction}

Well established magnetic chemically peculiar stars (also known as CP2 objects) 
for which no or only limited time-series photometric 
observations from the literature (Joshi et al.~\cite{Josh16})
are available, were selected from the catalogue of Ap, HgMn 
and Am stars by Renson \& Manfroid~(\cite{Rens09}). Although most known roAp stars are 
spectrally classified as SrCrEu, we extended our list of targets 
also to earlier spectral types, i.e. silicon stars. We have not used any 
other photometric criteria, such as, for example, Joshi et al.~(\cite{Josh16}) used 
within the Str{\"o}mgren-Crawford $uvby\beta$ system. In total, we 
selected 55 objects for observations. Two of these targets (HD 12098 and 
HD 137909) are already known roAp stars with very small photometric 
amplitude (HD 12098) and only spectroscopically detected variability (HD 
137909). From these 55 objects, 18 have already been observed before in 
order to detect roAp variability. However, since some roAp stars show 
strong variabilities in amplitude (e.g. White et al.~\cite{White11}; Medupe et al.~\cite{Medu15}), 
several observations in different nights are not a disadvantage.

The choice of photometric filter is also important in this respect. 
Medupe \& Kurtz~(\cite{Medu98}) showed that the amplitude of roAp variability 
significantly drops redwards of $\approx$ 4500\,\AA. Therefore, using a 
filter passband somewhere between the near-ultraviolet and the blue will 
be most suitable to detect roAp pulsations.

Our observations of roAp candidates were performed with several 
different telescopes, instruments and filters. We used the Johnson $B$ 
($\lambda_c$\,=\,4400\,\AA, FWHM\,=\,950\,\AA), SDSS $g'$ (4750, 1400) 
and Str{\"o}mgren $v$ (4100, 160) filters. Here is an overview of the 
basics characteristics of the telescopes and instruments.

\begin{enumerate}
\item Automatic Photoelectric Telescope (APT) T6 (0.75m) at Fairborn Observatory (Arizona, USA)
\item BOOTES-2, 0.6m RC telescope, EMCCD Andor iXon-889+ in Andaluc{\'i}a, Spain 
\item F(/Ph)otometric Robotic Atmospheric Monitor (FRAM-Auger) 0.3m FRAM-Auger Schmidt-Cassegrain, G4-16000 CCD
telescope of the Pierre Auger Observatory (Malarg{\"u}e, Argentina)
\item Masaryk University Observatory (MUO, Brno, Czech Republic), 0.6m, G2-4000 CCD 
\item Sandvreten Observatory (Uppsala, Sweden), 0.406m, f/9 RC telescope,
SBIG STL-6303 CCD
\item Mt. Suhora observatory (Poland)
\begin{itemize}
\item 0.034m refractor (MK), G2-402 CCD
\item 0.2m RC telescope (gs8in), SBIG ST-10 CCD
\item 0.6m Zeiss telescope, Apogee AltaU-47 CCD
\end{itemize}
\end{enumerate}

The integration times varied between 10 and 30\,s depending on the 
instrument and target brightness.

The photoelectric data of the APT were reduced by compensating for 
coincidence losses and subtracting sky background. Extinction 
corrections were determined with the standard Bouguer method from the 
target star data themselves.

The reductions of the time series CCD observations were performed using 
a standard technique. First, bias-subtraction, dark-correction and 
flat-fielding was applied. Then, differential aperture photometry was 
performed using two different program packages:
\begin{itemize}
\item Image Reduction and Analysis Facility (IRAF, Stetson~\cite{Stet87}),
\item C-Munipack under Windows\footnote{http://c-munipack.sourceforge.net/}.
\end{itemize}
To account for different seeing conditions, several apertures were used 
and tested. For each data set, we have chosen the aperture which yielded the 
smallest scatter in the differential target star light curves. 
If several comparison stars were available, 
these were checked individually to exclude variable objects.

The total reduced data set comprised 202 hours of observations in 59 
different nights (Table \ref{log}). The time series analysis of the 
differential light curves was conducted applying a standard Fourier 
technique and the Phase-Dispersion-Method. All computations were done 
within the programme package Peranso (Paunzen \& Vanmunster~\cite{Paun16}).

Defining the upper limit of variability is not straightforward and has often been discussed in the literature 
(Reegen et al.~\cite{Reegen08}). In general, the statistical significance of the noise in the Fourier spectrum is underestimated. 
We here employ a conservative approach and define the upper limit of variability as the upper envelope of the peaks 
in an amplitude spectrum (Fig. \ref{betaCrB}). 

\section{Analysis and Results} \label{analysis}

The time series analysis of the targets revealed only upper limits for 
variability in the frequency range of the known roAp stars. The upper 
limits range from 0.3 to 9.8\,mmag with a mean of 1.9\,mmag and a median 
of 1.3 mmag. This is well in the range of other ground based surveys 
(Handler \& Paunzen~\cite{Hand99}; Joshi et al.~\cite{Josh16}) with similar or larger telescopes. Although we 
have not detected any new roAp star, establishing the upper limits 
across the observed instability strip is also important 
for testing and improving models taking into account the excitation 
and visibility of pulsations under the influence of a magnetic field 
(Saio~\cite{Saio08}). In this respect, it would be also interesting to 
examine, for example, the high precision Kepler Space Telescope (Murphy~\cite{Murphy14}) or K2 
data for the absence of short-term variability among stars in this 
region of the Hertzsprung-Russell-diagram.

In the following, we want to discuss our results for the two already known
roAp stars in more detail.

{\it HD 12098:} This star was reported as a roAp pulsator with a period of 
about 7.7\,min by Girish et al.~(\cite{Giri01}). They presented photometric data in 
Johnson $B$ for 16 individual nights with durations of one to nine 
hours. The reported amplitudes range from 0.38 to 1.47\,mmag with a mean 
of 0.94\,mmag. Such a strong amplitude variation was also 
found, for example, for the roAp star HD 217522 (Medupe et al.~\cite{Medu15}) for which the 
amplitudes ranges from 2.0\,mmag to be hidden in the noise 
($<$\,0.3\,mmag). Another prominent case, among others, is HD 24355 (Holdsworth et al.~\cite{Holds16}).
This behaviour can be explained by the Oblique Pulsator Model 
(Kurtz~\cite{Kurt82}, Bigot \& Kurtz~\cite{Bigot11}, Saio et al.~\cite{Saio12}).
Within this model, the amplitude can even approach a zero value depending
on the inclination of the rotation axis to the line-of-sight as well as the magnetic to the 
rotation axis for a given rotation period. 
We observed HD 12098 in two consecutive nights and 
could not detect its variability in Str{\"o}mgren $v$ (Table \ref{log}), 
although our detection level should have permitted this. This case shows 
how important it is not only to have high quality observations during 
several nights but also to use a filter in the blue wavelength region 
(Medupe \& Kurtz~\cite{Medu98}).

\begin{figure}
\begin{center}
\includegraphics[width=80mm]{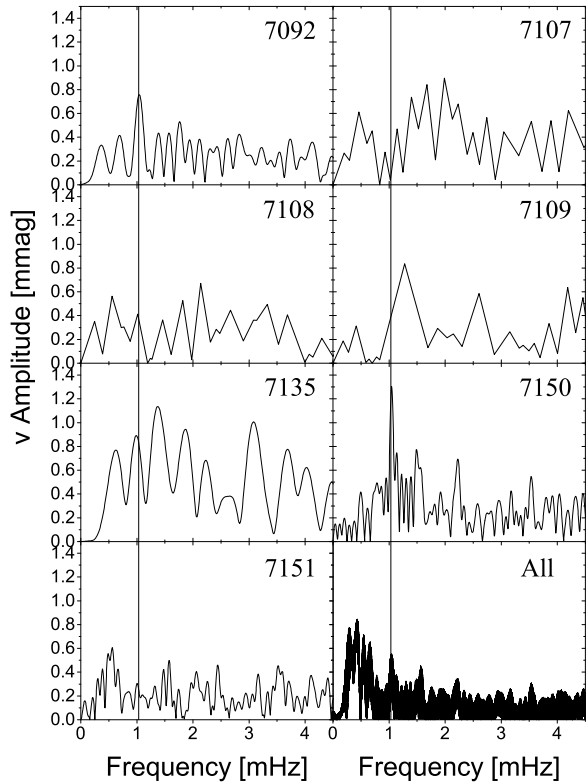}
\caption{The Fourier spectra of the observations for all nights (Table \ref{log}) of HD 137909 ($\beta$ CrB). 
The spectroscopically detected frequency of 1.03\,mHz is indicated by a vertical line.}
\label{betaCrB}
\end{center}
\end{figure}

{\it HD 137909 ($\beta$ CrB):} Hatzes et al.~(\cite{Hatzes04}) detected pulsational 
variability in their radial velocity measurements with a period of 
16.2\,min (1.029\,mHz) and an amplitude of about 140\,\ms. Later, 
Kurtz et al.~(\cite{Kurt07}) confirmed this period on the basis of individual lines of 
the first ionization stage of rare earth elements and in H$\alpha$. 
However, the amplitude was only 30\,\ms. Since no recent photometric 
time series were found in the literature, we observed HD 137909 in seven 
nights with the APT (Table \ref{log}). Although we cannot claim a 
statistically significant detection, we found in two nights a peak above 
the noise level very close (1.042(17) and 1.046(16)\,mHz, respectively) 
to to the spectroscopic period.
Within the errors, the photometric frequencies are 
in excellent agreement with the spectroscopic one. However, in the other 
five nights, no similar peaks are detected. In Fig. \ref{betaCrB}, the 
Fourier spectra of all nights are shown. Further photometric 
observations are needed to unambiguously confirm the spectroscopically found pulsational 
frequency. Again, the characteristics of the variability for HD 137909 
show the desire for precise photometric as well as spectroscopic 
observations over more than one night.

In general, it seems that the capabilities of ground-based photometric 
observations to detect new roAp stars have reached their limits. 
Suitable telescopes and instrumentations need to achieve a time 
resolution of less than one minute in the blue wavelength region 
(i.e. Johnson $B$ and Str{\"o}mgren $v$ filter) with a photometric accuracy in the 
sub-mmag region. At first 
sight, current on-going ground-based photometric surveys such as ASAS 
(Pojma{\'n}ski~\cite{Poj09} and references therein) and SuperWASP 
(Pollacco et al.~\cite{Pol06}), 
just to mention two, seem to neither have the time resolution nor the 
accuracy for this purpose (Bernhard et al.~\cite{bern15}; H{\"u}mmerich et al.~\cite{Huemm16}).

In this context, it is interesting to discuss the paper by 
Holdsworth et al.~(\cite{Holds14a}). These authors presented 350 stars which are variable on periods less
than 30\,min on the basis of SuperWASP data and concluded this 
variability is due to pulsation. The observations, using a broadband 
filter with a passband from 4000 to 7000\,\AA, are such that two 
consecutive 30\,s integrations at a given pointing are typically 
repeated every 10\,min. Such a time sampling is a-priori unfavourable 
for detecting variability in a period range up to 30\,min. This 
situation can however be mitigated by a significant number of 
observations over a long time base which then cover the whole 
pulsational phase. Among the 15 known roAp stars in their sample, Holdsworth et al.~(\cite{Holds14a}) were able to 
detect one (HD 12932 with a published amplitude of 4\,mmag in Johnson 
$B$). Other roAp stars of even higher amplitude, such as HD 99563 and HD 
101065, were not detected. In addition to the 
cadence of the data, saturation of objects brighter than 
$V$\,$<$\,8\,mag may also play a significant role here. Most of the known 
roAp stars are brighter than that limit and therefore their pulsations 
are very difficult to detect in SuperWASP data. This might explain the non-detection
of many known roAp stars.

Holdsworth et al.~(\cite{Holds14a}) also selected 543 Ap stars from Renson \& Manfroid~(\cite{Rens09}) on 
spectroscopic grounds and found no new roAp stars. This has to be taken 
into account when interpreting the results on the variability of the 
newly detected variables (as the authors themselves pointed out). In 
total, Holdsworth et al.~(\cite{Holds14a}) presented 10 out of the 350 variable objects as 
new roAp stars. Their reasoning that these are true roAp variables is 
based on classification resolution spectra and an enhancement 
of Sr, Cr and Eu, typical for CP2 stars (Ryabchikova \& Romanovskaya~\cite{Ryab17}). However, another 
observable that distinguishes roAp variables from other 
pulsators is the typical Pr-Nd anomaly (Ryabchikova et al.~\cite{Ryab04}) that should be 
searched for by high resolution spectroscopy. Apart from that, 
independent confirmation of variability of these and the remaining $> 
300$ stars and determination of its origin is desirable. Nevertheless, 
the capability of the SuperWASP data to detect new roAp stars on such a 
large scale is promising.

One way to provide a homogeneous study of the occurrence of the roAp 
phenomenon is the use of space-based data. Several roAp stars were found 
and analysed within the Kepler and K2 fields (Smalley et al.~\cite{Small15}) and references 
therein). Other roAp stars were investigated with 
BRITE-Constellation (Weiss et al.~\cite{Weiss16}), MOST (Gruberbauer et al.~\cite{Grub11}) and WIRE 
(Bruntt et al.~\cite{bruntt09}) satellite missions. The next milestone is expected to 
be the upcoming TESS mission (Ricker et al.~\cite{rick15}). With a cadence of 2\,min, a 
time-basis of up to one year and the predicted accuracy, it should be 
quite suitable to detect roAp stars in both hemispheres on a large 
scale. However, as TESS will observe in a very wide near infrared 
passband, it remains to be seen by how much the roAp pulsation 
amplitudes will be diminished in comparison to blue-filter data.
Nevertheless, the TESS data set will provide a homogeneous view on both,
noAp and roAp stars.

Another approach is the search for radial velocity shifts due to the 
pulsations (Ryabchikova et al.~\cite{Ryab07}; Kochukhov et al.~\cite{Koch13}). The amplitudes can reach up to 
5\,\kms\ with typical values between 100 and 500\,\ms\ (Elkin et al.~\cite{Elkin05}). 
Such accuracies are nowadays even achievable with a 60\,cm telescope 
(Pribulla et al.~\cite{Prib15}). However, the spectral and time resolution have to be 
appropriate for the expected pulsational period and line blending. But 
also here, the candidates have to be selected with care. Freyhammer et al.~(\cite{Frey08}), 
for example, observed nine luminous CP stars and found only upper limits 
for radial velocity amplitudes of 20 to 65\,\ms. The 350 photometric 
variable candidates reported by Holdsworth et al.~(\cite{Holds14a}) are excellent targets 
for such observations. Last, but not least, one can also search for the 
typical Pr-Nd anomaly (Ryabchikova et al.~\cite{Ryab04}) in known CP stars using high 
resolution spectroscopy, as a spectroscopic signature of the 
roAp phenomenon that appears to be an abundance differences of second and 
first ions of Pr and Nd of at least 1.5 dex and up to 2.5 dex. Stars 
with this spectral feature then need to be tested for rapid photometric 
variations.

\subsection{New Str{\"o}mgren-Crawford photometric criteria to search for roAp stars}

Although a significant number of stars were searched for roAp 
variability in the past, we want to estimate the amount of potential candidates 
not checked for pulsations to date. To this end, criteria to find such 
candidates have to be established. A basic selection can be done either 
by spectroscopic or photometric means (or by a combination of both). In 
the past, both approaches have been successfully applied 
(e.g. Handler \& Paunzen~\cite{Hand99}; Martinez et al.~\cite{Mart01}; 
Paunzen et al.~\cite{Paun12}).

For a spectroscopic-based selection, the catalogue by Renson \& Manfroid~(\cite{Rens09}) and 
its previous versions have been used. Since the roAp stars are normally 
characterized as cool CP SrCrEu stars, they can be traced in this 
catalogue or at least the CP1 (Am/Fm) stars can be excluded as 
candidates. However, this catalogue is an inhomogeneous compilation 
which is outdated. A new version is very much needed to perform an 
assessment of possible roAp candidates, not yet photometrically 
investigated.

For an efficient and viable selection based on photometry, 
systems are needed that provide parameters that are reddening 
free and are metallicity and luminosity sensitive.
For such a purpose, only intermediate to narrow 
band photometric systems are suitable. This excludes the usage of the 
2MASS (Skrutskie et al.~\cite{Skru06}) measurements, for example. The same is true for the 
SDSS photometric system used for APASS (Henden et al.~\cite{Hend15}). We mention just 
these two all-sky surveys because they (will) provide deep homogeneous 
photometric data from the UV to the NIR region. In principle, three 
well-known photometric systems are able to provide the required
information: Geneva 7-colour (Golay~\cite{Golay72}; Nicolet~\cite{Nico86}), 
Str{\"o}mgren-Crawford $uvby\beta$ (Str{\"o}mgren~\cite{Stro66}), and Vilnius
(Strai\v{z}ys \& Sviderskiene~\cite{StSv72}; Smriglio et al.~\cite{Smri86}).
Unfortunately, from these, only the 
Str{\"o}mgren-Crawford system is nowadays still relatively widely used. 
In addition, it was also used for the roAp candidate selection in the 
past (Joshi et al.~\cite{Josh09}).

In the literature, several definitions of the different photometric 
parameters within the Str{\"o}mgren-Crawford system are available. 
Because sometimes different definitions are given, we list the ones
used for our study, these are:

\begin{itemize}
	\item $\lbrack c_1 \rbrack$ = $c_1$ -- 0.19($b - y$)
	\item $\lbrack m_1 \rbrack$ = $m_1$ + 0.33($b - y$)
	\item $\lbrack u - b \rbrack$ = ($u - b$) -- 1.53($b - y$)
  \item $a0$ = 1.54$\lbrack m_1 \rbrack$ + 0.74$r$ -- 0.27
  \item $r$ = 0.35$\lbrack c_1 \rbrack$ -- $\beta$ + 2.565
\end{itemize}

It is important to use the whole variety of these indices because
they are sensitive to different parameters as briefly discussed in the following. 
All known roAp stars, 
besides one (KIC~7582608 with $T_\mathrm{eff}$\,$=$\,8700\,K; 
Holdsworth et al.~\cite{Holds14b}), fall in the ``cool region'' of 
$T_\mathrm{eff}$\,$\leq$\,8500\,K as defined by Napiwotzki et al.~(\cite{Napiw93}). There, 
$\beta$ and $\lbrack m_1 \rbrack$ are measures for $T_\mathrm{eff}$ 
and $\lbrack c_1 \rbrack$ an indicator for the Balmer discontinuity, 
thus the luminosity; $\lbrack m_1 \rbrack$ is also sensitive to 
metallicity. For the intermediate region (11000\,K\,$\leq$\,
$T_\mathrm{eff}$\,$\leq$\,8500\,K), these indices are replaced by $a0$ 
and $r$. The $\lbrack u - b \rbrack$ colour index is sensitive to 
$T_\mathrm{eff}$ throughout the whole range and was also used to 
calibrate CP stars (M{\'e}gessier~\cite{Meges88}). Although there are three different 
indices which are sensitive to $T_\mathrm{eff}$, this is vital for the 
study of CP stars because they exhibit significant flux depressions at 
4100, 5200, and 6300\,\AA\ (Adelman~\cite{Adel80}). Measuring $T_\mathrm{eff}$ 
using different spectral regions avoids the introduction of systematics
in the analysis.

\begin{figure}
\begin{center}
\includegraphics[width=80mm]{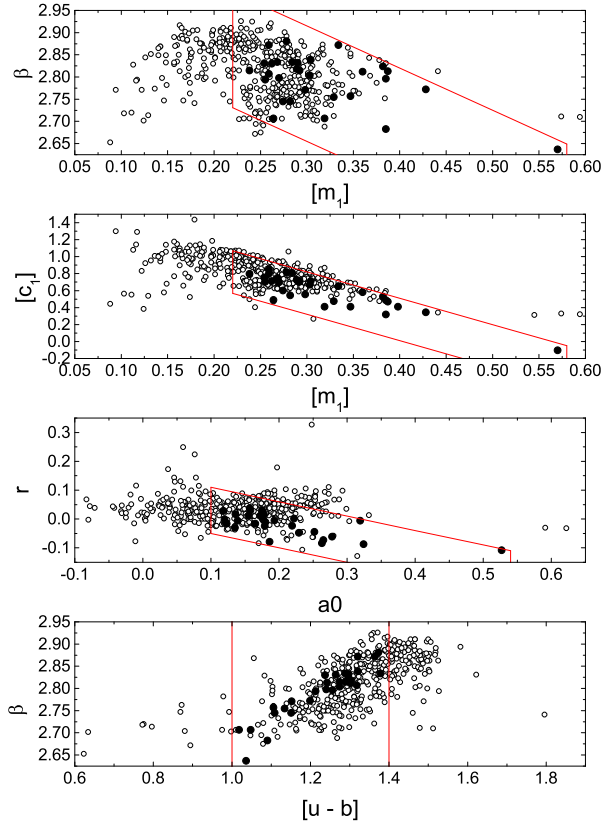}
\caption[]{The location of 416 noAp (open circles) and 42 roAp (filled circles) within
three reddening free Str{\"o}mgren-Crawford $uvby\beta$ diagrams. The red lines denote
the boxes defined from the known roAp stars in which we searched for new candidates.}
\label{uvbybeta}
\end{center}
\end{figure}

\begin{table*}
\begin{center}
\caption[]{The list of 4646 new roAp candidates selected on the basis of Str{\"o}mgren-Crawford
$uvby\beta$ diagrams. The errors are taken from Paunzen~(\cite{Paun15a}).}
\label{candidates_list}
\begin{tabular}{cccccccccccccccc}
\hline\noalign{\smallskip}
TYC & $V$ & $\sigma V$ & $N$ & $(b-y)$ & $\sigma(b-y)$ & $N$ & $\lbrack m_1 \rbrack$ & $\sigma\lbrack m_1 \rbrack$
& $N$ & $\lbrack c_1 \rbrack$ & $\sigma\lbrack c_1 \rbrack$ & $N$ & $\beta$ & $\sigma\beta$ & $N$ \\ 
\hline\noalign{\smallskip}
5-334-1 & 9.000 &   & 1 & 0.208 &   & 1 & 0.169 &   & 1 & 0.740 &   & 1 & 2.732 &   & 1 \\
5-958-1 & 8.430 &   & 1 & 0.173 &   & 1 & 0.242 &   & 1 & 0.742 &   & 1 & 2.779 &   & 1 \\
11-1234-1 & 8.014 & 0.013 & 4 & 0.229 & 0.003 & 4 & 0.164 & 0.006 & 4 & 0.698 & 0.013 & 4 & 2.719 & 0.003 & 2 \\
13-974-1 & 7.663 & 0.024 & 2 & 0.104 & 0.001 & 2 & 0.211 & 0.001 & 2 & 0.914 & 0.008 & 2 & 2.847 &   & 1 \\
19-1437-1 & 6.098 & 0.016 & 3 & 0.154 & 0.005 & 3 & 0.194 & 0.007 & 3 & 0.805 & 0.021 & 3 & 2.796 & 0.007 & 3 \\
20-386-1 & 8.122 & 0.004 & 2 & 0.179 & 0.001 & 2 & 0.196 & 0.004 & 2 & 0.799 & 0.015 & 2 & 2.774 &   & 1 \\
23-173-1 & 7.590 &   & 1 & 0.183 &   & 1 & 0.176 &   & 1 & 0.798 &   & 1 & 2.750 &   & 1 \\
23-524-1 & 9.650 &   & 1 & 0.187 &   & 1 & 0.199 &   & 1 & 0.732 &   & 1 & 2.762 &   & 1 \\
25-677-1 & 8.550 &   & 1 & 0.170 &   & 1 & 0.192 &   & 1 & 0.777 &   & 1 & 2.762 &   & 1 \\
27-637-1 & 8.370 &   & 1 & 0.181 &   & 1 & 0.194 &   & 1 & 0.708 &   & 1 & 2.739 &   & 1 \\
31-919-1 & 8.666 &   & 1 & 0.172 &   & 1 & 0.235 &   & 1 & 0.804 &   & 1 & 2.806 &   & 1 \\
32-127-1 & 8.390 &   & 1 & 0.199 &   & 1 & 0.183 &   & 1 & 0.720 &   & 1 & 2.754 &   & 1 \\
32-385-1 & 8.860 &   & 1 & 0.202 &   & 1 & 0.178 &   & 1 & 0.687 &   & 1 & 2.721 &   & 1 \\
35-339-1 & 9.360 &   & 1 & 0.188 &   & 1 & 0.171 &   & 1 & 0.725 &   & 1 & 2.750 &   & 1 \\
41-782-1 & 8.257 & 0.011 & 2 & 0.194 & 0.001 & 2 & 0.171 & 0.007 & 2 & 0.674 & 0.008 & 2 & 2.745 &   & 1 \\
41-820-1 & 8.300 &   & 1 & 0.168 &   & 1 & 0.182 &   & 1 & 0.771 &   & 1 & 2.759 &   & 1 \\
\noalign{\smallskip}\hline
\multicolumn{16}{l}{Only a portion of this table is shown here to demonstrate its form and content. A machine-readable version of the full table is available.}
\end{tabular}
\end{center}
\end{table*}

Often, the $\delta c_1$ and $\delta m_1$ indices were applied to select 
targets (Joshi et al.~\cite{Josh06}). Both indices are defined as the deviation from a 
standard line of unreddened, solar abundant, main-sequence stars such
that they are negative for stars with overbundances of elements. 
One has to keep in mind that $\delta c_1$ is also sensitive to 
luminosity. The standard line can be defined either by observational 
data or by stellar atmosphere models and synthetic photometry. Thus, the 
calculation of $\delta c_1$ and $\delta m_1$ is not straightforward and 
the interpretation of these indices can be ambiguous.

We now evaluate how the regions of known noAp and roAp stars overlap in 
the different photometric Str{\"o}mgren-Crawford diagrams. For this, 
lists for both groups are needed.

We have searched the literature for stars which were investigated for 
roAp variability. Of course, these investigations were performed with 
different instruments and filters. Also the overall quality 
of the observing nights, the time base, and cadence play a role. It is 
therefore not really possible to compare directly all these different 
sources when it comes to define to which limit a star does not show any 
roAp variability (see also the discussion above about amplitude 
variations over time). We used the following 21 references to generate a 
list of noAp objects 
\begin{itemize} 
\item Kurtz~(\cite{Kurt84})
\item Matthews \& Wehlau~(\cite{Matt85}) 
\item Heller \& Kramer~(\cite{Heller88}) 
\item Matthews et al.~(\cite{Matt88}) 
\item Belmonte~(\cite{Belm89}) 
\item Kurtz~(\cite{Kurt89}) 
\item Martinez~(\cite{Mart89}) 
\item Weiss \& Schneider~(\cite{Weiss89})
\item Kreidl~(\cite{Kreidl91}) 
\item Nelson \& Kreidl~(\cite{Nelson93}) 
\item Schutt~(\cite{Schutt93}) 
\item Martinez \& Kurtz~(~\cite{Mart94}) 
\item Dorokhova \& Dorokhov~(\cite{Dorok98}) 
\item Handler \& Paunzen~(\cite{Hand99}) 
\item Martinez et al.(~\cite{Mart01}) 
\item Joshi et al.~(\cite{Josh06}) 
\item Joshi et al.~(\cite{Josh09}) 
\item Paunzen et al.~(\cite{Paun12}) 
\item Paunzen et al.~(\cite{Paun13}) 
\item Paunzen et al.~(\cite{Paun15b}) 
\item Joshi et al.~(\cite{Josh16}) 
\end{itemize} 
In total, the final list (available in 
electronic form) includes 620 stars from which 109 objects have been 
found in more than one reference. For the next step, the 61 roAp stars listed 
in Joshi et al.~(\cite{Josh16}) were taken and all objects searched for $uvby\beta$ 
colors in the catalogue by Paunzen~(\cite{Paun15a}). This catalogue includes only 
stars with Tycho identifiers and is therefore limited to objects 
brighter than 15th magnitude (98\% of the stars in that catalogue
are between 4th and 12th 
magnitude). In the following, only stars with complete $uvby\beta$ 
measurements can and will be considered. Those are 416 (67\% of 
all noAp) and 42 (69\% of all roAp) objects, i.e. a representative 
number for both subgroups.

In Fig. \ref{uvbybeta} we present three diagrams of unreddened indices: 
$\beta$ versus $\lbrack m_1 \rbrack$, $\lbrack c_1 \rbrack$ versus 
$\lbrack m_1 \rbrack$ and $r$ versus $a0$. It is obvious that the blue 
boundary of the roAp phenomenon is well defined and a significant number 
of objects hotter than the known roAp stars have already been 
observed. However, the red border appears more fuzzy. 
The reddest star in these diagrams is HD 101065 (Przybylski's Star), but 
also HD 69013 and HD 143487 are located in the region with $a0 > 0.3$. 
These stars also have the largest $\lbrack m_1 \rbrack$ and $\lbrack u - b \rbrack$ 
values.
For HD 101065, several detailed investigations are available in the literature. 
Shulyak et al.~(\cite{Shulyak10}) concluded that there is nothing special in HD 101065 
compared to other known cool CP and roAp stars except its relatively low 
$T_\mathrm{eff}$ (6\,400\,K) and a high abundance of rare earth 
elements. Erroneous measurements of $uvby\beta$ can also be excluded as 
a cause for the outstanding location in the diagrams because there are 
four consistent different references over almost two decades available. 
Therefore, the range for the search of roAp candidates on the basis of 
Str{\"o}mgren-Crawford $uvby\beta$ colours should be significantly 
extended to cooler CP stars. In this respect, it is especially 
interesting to study the possible overlap with solar-type oscillations 
which are in the same frequency region, but according to theory 
should be suppressed by the effects of the magnetic field that are 
required to excite roAp pulsations (Balmforth et al.~\cite{balmforth01}). 
Helioseismology found that the mode amplitudes of solar acoustic
oscillations are anti-correlated with solar
magnetic activity. 

Using the diagrams shown in Fig. \ref{uvbybeta}, we define new and 
extended boundaries for the location of known roAp stars for 
a further search of other variable stars of this type:

\begin{eqnarray*}
	0.58 > &\lbrack m_1 \rbrack& > 0.22 \\
	1.40 > &\lbrack u - b \rbrack& > 1.00 \\
	0.54 > &a0& > 0.10 \\
	\lbrack 3.2 - (0.95*\lbrack m_1 \rbrack)] > &\beta& > [2.94 - (0.95*\lbrack m_1 \rbrack)] \\
	\lbrack 1.75 - (3.1*\lbrack m_1 \rbrack)] > &\lbrack c_1 \rbrack& > [1.25 - (3.1*\lbrack m_1 \rbrack)] \\
%	&r& < 0.05 \\
	\lbrack 0.16 - (0.5*a0)] > &r& > 0.5*a0
\end{eqnarray*}

Applying these relations to the catalogue by Paunzen~(\cite{Paun15a}), we find a 
list (available in electronic form) of 4646 new candidates (Table \ref{candidates_list}). This sample 
is intentionally a-priori unbiased by any additional spectral type 
information, and can be used as a starting point for searches for roAp variables.

Criteria for selecting roAp candidates from Str{\"o}mgren-Crawford 
photometry can be manifold. For instance, comparing with the boundaries 
of the roAp phenomenon as defined by Joshi et al.~(\cite{Josh16}), who use different 
parameters (and exclude HD 101065), more stars satisfy our criteria. The 
main reasons are that we do allow for measurement errors, and for a 
possible extension of the borders of the roAp domain compared to 
current knowledge. Ideally, one would desire an inspection of all 
stars from our sample to define (non)variability. Such a large sample 
can only be comprehensively investigated with data already collected by automatic surveys or
by forthcoming space photometry missions.

\section{Conclusions}

We investigated some characteristics of noAp and roAp stars using the 
available data from the literature and new photometric observations. 
These stars are interesting because the driving mechanism of their 
oscillation modes is most probably due to the ``classical'' 
$\kappa$-mechanism operating in the hydrogen ionisation zone, with 
some possible contributions by turbulent pressure (Cunha et al.~\cite{cunha13}). The 
roAp stars possess a structured and stable magnetic field which triggers 
diffusion and stratification of chemical elements. The latter leads 
to strong chemical peculiarities on the stellar surface (spots) which 
give the possibility to directly measure the rotational period due to 
light variations. Such a combination is unique among the stars of the 
upper main sequence where convection does not dominate the outer stellar 
atmosphere.

An enormous effort was spent in the last three decades to search for 
roAp stars. The detection of these variables is made difficult by the 
low pulsation amplitudes (typically only a few mmag) that are largest 
in the blue optical spectral region and the excited period range (five 
to twenty five minutes). Other complicating facts are that the amplitudes 
can vary strongly even between consecutive nights, and that the amplitudes 
can be so low (or even zero according to the Oblique Pulsator Model) that they are 
undetectable from the ground. This also 
makes the definition of a noAp star, i.e. an object that does not 
pulsate, difficult; one always has to include the upper limit for a given 
time basis and filter.

The results of our new observations of 55 stars (202 hours in 59 
different nights) were combined with already published ones for apparent 
noAp and roAp stars. For the objects with available 
Str{\"o}mgren-Crawford $uvby\beta$ photometry, reddening-free diagrams 
were generated and analysed. From these diagrams we concluded that the 
blue border of the observed roAp instability strip is well determined 
and investigated. The opposite is true for the red border where still 
only a small number of possible candidates have been observed.
On the 
basis of these diagrams, we defined limits for $\lbrack m_1 \rbrack$, 
$\lbrack c_1 \rbrack$, $\lbrack u - b \rbrack$, $a0$, $r$ and $\beta$ in 
which we searched for new roAp candidates. In total, we found 4646 
stars brighter than 15th magnitude. We suggest to use surveys like 
SuperWASP or the forthcoming TESS mission to search for pulsation in 
these stars.

On the basis of the published and newly observed data for HD 12098 and 
HD 137909, we discussed the necessity of high quality photometric 
observations in several nights to establish variability for this star 
group. This difficulty may be overcome by searching for radial velocity 
variability or the unique Pr-Nd anomaly present in roAp stars, followed 
by a photometric search.

Further ground and space based efforts are very much needed to shed more 
light on the roAp phenomenon. In particular, an analysis of the possible 
connection with solar-type oscillations is necessary to understand the 
changes from radiative to convective dominated stellar atmospheres and 
their different/common pulsational behaviour in the presence of a 
magnetic field.

\begin{acknowledgements}
This project was supported by the grant 7AMB14AT030 (M\v{S}MT). We would 
like to thank the Pierre Auger Collaboration for the use of its 
facilities. The operation of the robotic telescope FRAM was supported by 
the EU grant GLORIA (No. 283783 in FP7-Capacities program) and by the 
grant of the Ministry of Education of the Czech Republic 
(M\v{S}MT-\v{C}R LM2016038). The data calibration and analysis related 
to the FRAM telescope is supported by the Ministry of Education of the 
Czech Republic M\v{S}MT-\v{C}R LG15014 and EU-M\v{S}MT 
CZ.02.1.01/0.0/0.0/16\_013/0001402. The operation of the Fairborn APT is 
secured by the Polish NCN grant 2015/18/A/ST9/00578.
\end{acknowledgements}

\label{lastpage}
\end{document}